\documentstyle[aps,prl,epsfig]{revtex}
\twocolumn

\title{Frobenius-Perron Resonances for Maps with a Mixed Phase Space}
\author{Joachim Weber$^1$, Fritz Haake$^1$, and Petr \v{S}eba$^2$}
\address{$1$ Fachbereich Physik, Universit\"at-GH Essen, 45117 Essen,
Germany}
\address{$2$ Institute of Physics, Czech Academy of Sciences, Prague, Czech
Republic}

\begin{document}

\draft
\date{Date: \today}
\maketitle

\begin{abstract}
Resonances of the time evolution (Frobenius-Perron)
operator ${\cal P}$ for phase space densities have
recently been shown to play a key role for the
interrelations of classical, semiclassical and
quantum dynamics. Efficient methods to determine
resonances are thus in demand, in particular for
Hamiltonian systems displaying a mix of chaotic
and regular behavior. We present a powerful method
based on truncating ${\cal P}$ to a finite matrix
which not only allows to identify resonances but
also the associated phase space structures. It is
demonstrated to work well for a prototypical
dynamical system.
\end{abstract}

\begin{figure}
\epsfxsize=0.9\textwidth
\epsffile{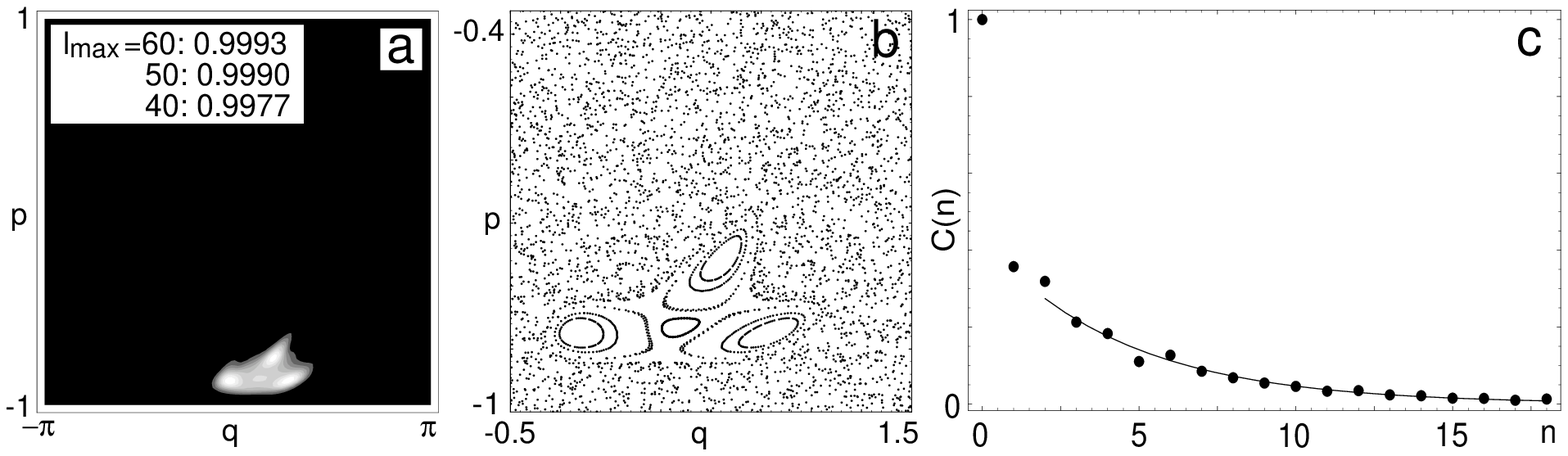}
\caption{(a) The eigenfunction to the almost unimodular eigenvalue $0.9993$
for $\tau
= 4, l_{max}=60$  is sharply
localized on elliptic islands surrounding a period-$3$-orbit in phase space.
(b) Phase space portrait of the elliptic islands supporting the
eigenfunction.
(c) The decay of the correlator $C(n)$ (dots) [with the initial density  
localized in the region shown in figure \protect\ref{figure2}c]  and the  
decay  predicted by the corresponding resonance $0.8103$ (full line) agree  
well (see also table \protect\ref{table2}).}
\label{figure1}
\end{figure}

Effectively irreversible behavior of classical Hamiltonian systems can be
elucidated by studying the phase
space density and its propagator, the  Frobenius-Perron operator ${\cal P}$.
Due to Liouville's theorem
${\cal P}$ can be represented by an infinite unitary matrix whose spectrum
lies on the unit circle in the
complex plane. Nevertheless, means and correlation functions of observables
can relax (see figure \ref{figure1}c)
with damping factors known as (Ruelle-Pollicott) resonances
\cite{Ruelle,Ruelle2,Pollicott,Baladi} of  ${\cal P}$.
These resonances have recently attracted attention, e.g. in a superanalytic
approach to uni\-versal
fluctuations in quantum (quasi-) energy spectra which originated from the
physics of disordered systems.
In that approach the Frobenius-Perron resonances constitute a link
between classical and quantum chaos \cite{Altshuler,Zirnbauer}. There is
even a recent experiment where
quantum fingerprints of Ruelle-Pollicott resonances are identified
\cite{PanceLuSridhar}.
To further clarify the interrelations between classical, semiclassical, and
quantum behavior, a practical
scheme to actually determine classical resonances is called for which is
free of restrictions of previous
investigations, like hyperbolicity, one-dimensional (quasi-) phase space or
isolation of the phase-space
regions causing intermittency.

\begin{figure}
\epsfxsize=0.9\textwidth
\epsffile{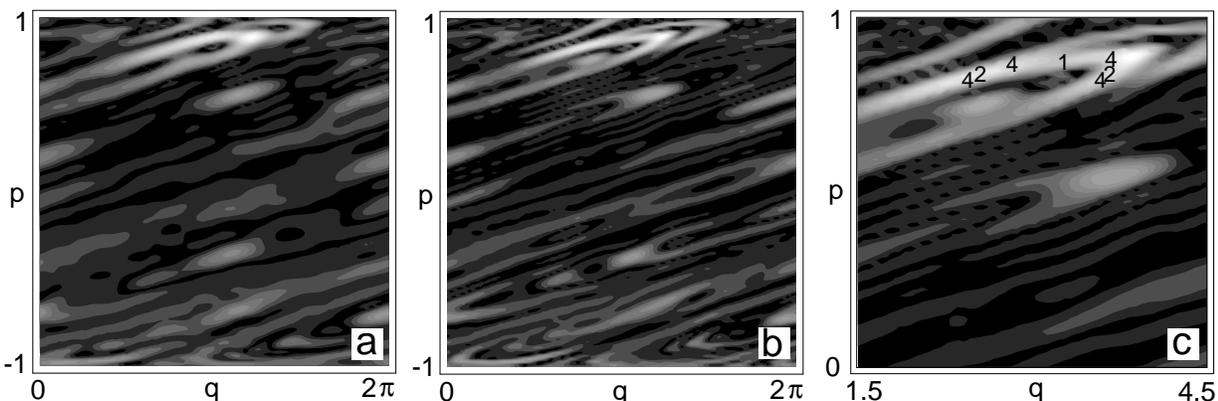}
\caption{Eigenfunction to the resonance-eigenvalue $0.81$ for $\tau = 10$
and
(a) $l_{max} = 40$, (b) $l_{max} = 60$. While the coarse structures are
identical,
finer structures appear in the eigenfunction with the higher resolution.
(c) Magnification of the region of large amplitude in figure
\protect\ref{figure2}b.
The numbers indicate the positions of the three periodic orbits of lengths
$1,2$
and $4$  which are used in the cycle expansion up to order $n_{max} = 4$.
}
\label{figure2}
\end{figure}

Our thus motivated quantitative investigations into Hamiltonian
systems with mixed phase space, a still
largely unexplored area of great interest and promise, lead us to the
discrete unimodular Frobenius-Perron eigenvalues
(with eigenfunctions localized in islands of regular motion around elliptic
periodic orbits, see figure
\ref{figure1}a) and to resonances, smaller than unity in modulus (with
eigenfunctions  localized
on the unstable manifolds of hyperbolic periodic orbits, see figure
\ref{figure2}a-c).
Both the discrete spectrum and the resonances are determined by
diagonalizing  truncated
Frobenius-Perron matrices ${\cal P}^{(N)}$ and studying the cut-off
dependence of its eigenvalues.
Using the information about eigenfunctions, we then reproduce resonances by
the so-called cycle expansion of
periodic-orbit theory and furthermore through decay rates of correlation
functions.
We should add that similarly motivated but technically different (not
involving eigenfunctions and employing
external noise) efforts to determine resonances can be found in
Refs.~\cite{Fishman0,Fishman1,Fishman2}.

As a prototypical dynamical system we have employed the kicked top,
i.e. a periodically kicked angular momentum $\vec{J} = (j \sin \theta \cos
\varphi, j \sin \theta \sin \varphi, j \cos \theta)$ of conserved length $j$
whose
phase space is the sphere $\vec{J}^2/j^2 = 1$; we confront a single degree
of freedom with the ``azimutal'' angle $\phi$ as the coordinate and the
cosine
of the ``polar'' angle $\theta$ the conjugate momentum.
The dynamics is specified as a stroboscopic area preserving
map $M$ on phase space.
It consists of rotations $R_z(\beta_z), R_y(\beta_y)$  about the $y-$ and
$z-$axes and a ``torsion'', i.e. a nonlinear
rotation $R_z(\tau \cos \theta)$ about the $z-$axis which changes $\varphi$
by $\tau \cos \theta$,
\begin{equation}
M=R_z(\tau \cos \theta) R_z(\beta_z) R_y(\beta_y) \, .
\end{equation}
The equivalent map of the phase space density $\rho$ is brought about by the  
Frobenius-Perron operator ${\cal P}$,
\begin{equation}
{\cal P} \rho(\cos\theta, \varphi) = \rho(M^{-1}(\cos\theta, \varphi)) \, .
\end{equation}
We keep $\beta_z = \beta_y = 1$ fixed and vary the torsion constant $\tau$,
starting with the integrable case $\tau=0$.
Increasing values of $\tau$ bring about more and more
chaos until
for $\tau > 10$ elliptic islands have become so small that they are
difficult to detect.
We shall focus on $\tau = 4$ (roughly $90\%$ of the
phase space dominated by chaos) and $\tau=10$ (more than $99\%$ chaos).

A Hilbert space of phase space functions on the sphere is spanned by the
spherical
harmonics $Y_{lm}(\theta, \varphi)$ with $l=0, 1, 2, \dots$ and $|m|\le l$.
These functions are
ordered with respect to phase space resolution by the index $l$: if all
$Y_{lm}$
with $0\le l\le l_{max}$ are admitted phase space structures of area roughly
$4 \pi/(l_{max}+1)^2 $
can be resolved. If we so truncate the infinite Frobenius-Perron matrix
${\cal P}_{lm,l'm'}$, we
(i) destroy unitarity, (ii) restrict the spectrum to $N=(l_{max}+1)^2$
discrete
eigenvalues whose moduli cannot exceed unity, and (iii) renounce the
resolution
of phase space structures of linear dimension below
$\sqrt{4\pi}/(l_{max}+1)$.
Upon diagonalizing the truncated $N\times N\,$matrix ${\cal P}^{(N)}$ and
increasing $N$ we find the ``newly born'' eigenvalues close to the origin
while the ``older'' ones move about in the complex plane. ``Very old'' ones
eventually
settle for good. If the classical dynamics is integrable ($\tau = 0$ or
$\beta_y=0$),
the asymptotic large$-N$ loci are back to the unit circle, where the full
${\cal P}$ has
its spectrum. But not so for a mixed phase space: while some eigenvalues of
${\cal P}^{(N)}$ ``freeze'' with unit moduli, others come to rest inside the
unit
circle as $N \to \infty$. Table \ref{table1}
illustrates how non-unimodular  eigenvalues found for the kicked top with
$\tau = 10$ at
$l_{max}= 40$ remain in their positions  as  $l_{max}$ is increased to
$l_{max} = 50, 60$
and $70$.

We could pass over such findings and speak of the danger of tampering with
infinity,
were there not good reasons for and a physical interpretation of the
existence of
such stable non-unimodular eigenvalues.

The following qualitative argument suggests the persistence of
non-unimodular
eigenvalues as $N\to\infty$ for non-integrable dynamics. In contrast to
regular motion, chaos
brings about a hierarchy of phase space structures which extends without end
to ever finer
scales. A truncated Frobenius-Perron operator ${\cal P}^{(N)}$ must reflect
the flow of probability
towards the unresolved scales as a loss, however large the cut-off $N$ may
be chosen.

Arguments from perturbation theory \cite{Titchmarsh,Reed} indicate that any
non-unitary
approximation to a unitary operator with continuous spectrum  has some
eigenvalues in positions
near (non-unimodular) resonances of the unitary operator, i.e. poles of the
resolvent in a
higher Riemannian sheet. The perturbation series for such an eigenvalue does
not
converge but produces, with increasing order, a sequence of points
concentrated in the
neighborhood of the respective resonance.
It is intuitive to interpret the freezing of non-unimodular eigenvalues
(which need not be
a strict convergence) as analogous to the ``spectral concentration'' known
from
perturbation theory.

To find further evidence for our interpretation of frozen  eigenvalues as
resonances we have looked at the eigenfunctions of ${\cal P}^{(N)}$, with
the following
salient results. Eigenvalues freezing with unit moduli have eigenfunctions
located on elliptic
islands of regular motion surrounding elliptic periodic orbits in phase
space.
Such islands are bounded by invariant tori which form impenetrable barriers
in
phase space. We can thus expect the function constant inside the elliptic
islands around a
$p-$periodic orbit and zero outside to be an eigenfunction of ${\cal P}$
with eigenvalue
unity. Similarly reasoning we expect, for $p>1$, the $p$th roots of unity to
arise as eigenvalues
as well; their eigenfunctions should have constant moduli and be invariant
under ${\cal P}^p$.
For the kicked top with $\tau = 4$ and $l_{max} = 60$ an eigenfunction with
the eigenvalue $0.9993$, i.~e.
almost at unity, is shown in figure \ref{figure1}a. It is localized on the
three islands around an elliptic
orbit of period three (see figure \ref{figure1}b) and does have the two
expected partners.
We have indeed found frozen eigenvalues near the $p$-th roots of unity and
their
eigenfunctions localized near elliptic period-$p$ orbits for $p$ up to 6;
without much further
effort such signatures of higher periods could be identified.

Now on to the eigenvalues freezing with moduli smaller than unity.
Once such freezing has been observed the corresponding eigenfunction has
approached its final shape on the resolved phase space scales.
The eigenfunctions are sharply localized around unstable manifolds of
hyperbolic periodic
orbits, ones  with low periods at first since these are easiest to resolve;
but with growing
$l_{max}$ more complex orbits of higher periods appear in the ``support'' of
eigenfunctions.
Even though all periodic orbits contributing to the structure of an
eigenfunction have similar
stability coefficients and even though the latter do describe the rate of
mutual departure of
neighboring trajectories it would  be too naive to simply identify
resonances with stability
coefficients; we shall rather have to resort to cycle expansions further
below.

Just as for the eigenvalues there is no strict convergence of the
eigenfunctions.
With increasing resolution new structures on finer scales become visible, in
correspondence
with the infinitely convoluted shape of the unstable manifolds (see figure
\ref{figure2}a-b).
Since no finite approximation ${\cal P}^{(N)}$ accounts for arbitrarily fine
structures one encounters the aforementioned loss of probability from
resolved to unresolved
scales. Not even in the limit $N\to\infty$ can the unitarity of ${\cal P}$
be restored: rather,
the eigenfunctions tend to singular objects outside the Hilbert space, in
tune with a
continuous spectrum  of $\cal P$.

The reader may have noticed that all eigenvalues in table \ref{table1}
are real or almost imaginary. In fact, all eigenvalues we have identified as
frozen inside the unit circle have phases corresponding to those of roots of
unity, a fact
demanding explanation. Clearly, since ${\cal P}^{(N)}$ is real the
eigenvalues are either real
or come in complex conjugate pairs, but no other phase than zero is
distinguished by that
argument. Again, the eigenfunctions offer further clues. We find that
the phases of the complex eigenvalues are determined by the length $p$ of
the
shortest periodic orbit present in an eigenfunction $f$ as those of the
$p$-th roots of unity.
The following intuitive argument indicates that this is to be expected.

Assume an eigenfunction is mostly concentrated around a shortest unstable
orbit
with period $p$ as well as a longer one with period $p'$. Denote by
$\delta_{p,n}$ a
``characteristic function'' which is constant near the $n$-th point of the
period-$p$ orbit,
$n=1\ldots p$, and zero elsewhere. The truncated Frobenius-Perron operator
${\cal P}^{(N)}$ maps
$\delta_{p,n}$ into ${\cal P}^{(N)}\delta_{p,n}=r_p\delta_{p,n+1}$ with the
real positive factor
$r_p$ smaller than unity accounting for losses, in particular to unresolved
scales. Independent
linear combinations of the $\delta_{p,n}$ can be formed as
$f_{pk}=\sum_{n=1}^p{\rm e}^{{\rm i}2\pi k n/p}\delta_{p,n}$ with
$k=1\ldots p$. Now consider a sum of two such functions, $g=f_{pk}+
f_{p'k'}$,
and apply ${\cal P}^{(N)}$. For $g$ to qualify as an approximate
eigenfunction we must
obviously have $r_p\approx r_{p'}$ and $k/p=k'/p'$. But then indeed ${\cal
P}^{(N)}g\approx
r_p{\rm e}^{{\rm i}2\pi k/p}g$ and $[{\cal P}^{(N)}]^pg\approx r_p^pg$. The
phase is thus dictated
by the shortest orbit. Needless to say, the argument is identical to the one
used before for the
eigenfunctions living in islands around elliptic orbits, save for $r_p=1$ in
those regular
cases.  Since orbits of low period are most likely to be resolved first, the
eigenvalues
found for $l_{max}= 40$ in table \ref{table1} have phases according to
$p=1,2$ and $4$.

Knowing which orbits are linked to a non-unimodular eigenvalue, we can adopt
a cycle expansion
to calculate decay rates from periodic orbits \cite{Cvitanovic}.
A cycle expansion of the spectral determinant, i.e. the characteristic
polynomial of the Frobenius-Perron operator, allows for the calculation of
resonances in hyperbolic system with high accuracy \cite{Christiansen}.
The spectral determinant is expressed in terms of the traces of the
Frobenius-Perron operator ${\rm Tr}\;{\cal P}^{n}$ as
$d(z) = \prod_{n=1}^{\infty} \exp\left(-\frac{z^n}{n} {\rm Tr}\; {\cal P}^n
\right)$
and subsequently expanded as a finite polynomial up to some order $n_{max}$.
Only the first $n_{max}$  traces are required for the calculation of this
polynomial.
The traces  ${\rm Tr}\; {\cal P}^n$ are calculated by summing over
hyperbolic
periodic orbits of length $n$ as ${\rm Tr}\; {\cal
P}^n=\sum\frac{1}{|\det(1-J)|}$ where
the $2\times 2$ matrix $J=\partial M^n(X)/\partial X$
is the linearized map $M^n$ evaluated at any of the points of a contributing
period-$n$ orbit and $X=(\cos\theta, \varphi)$ the phase space point.
The zeros of the polynomial which are insensitive against an increase of
$n_{max}$ are inverses of resonances.

The condition under which the ordinary cycle expansion of a spectral
determinant converges is that all periodic orbits are hyperbolic and
sufficiently unstable \cite{Ruelle2,Cvitanovic,Christiansen}.
But if we only consider one ergodic region in phase space at a time, i.e.
bar contributions from elliptic orbits, and impose a stability bound by
including only the relatively few hyperbolic orbits identified in an
eigenfunction of ${\cal P}^{(N)}$, we can still use the cycle expansion as
follows.We assume the spectral determinant factorized as
$d(z) = \prod_{i=1}^\infty  d_i(z)$ with
one factor $d_i$ for the family of eigenfunctions to which a given set of
periodic orbits contributes.
Each such factor $d_i(z)$ is then  calculated separately with the above well
known expressions
but restricting the periodic-orbit sum for the trace ${\rm Tr}\; {\cal P}^n$
to the orbits previously
identified as contributing to the eigenfunctions.  In table \ref{table2}
resonances reproduced
via the spectral determinant from only few orbits show surprisingly good
agreement with the
resonance-eigenvalues of ${\cal P}^{(N)}$  for $\tau = 10$ and $l_{max}=60$.
The index $n_{max}$ gives the order up to which the spectral determinant is
expanded, i.e. the
length of the longest (pseudo-)orbits employed. The total number of orbits
used is given in brackets
behind the resonances. For the resonance $0.8103$ the three relevant orbits
of period $1,2$ and $4$
are marked in the magnified region of the eigenfunction ($l_{max}=60$) in
figure \ref{figure2}c.
The first repetition of the single period-2-orbit contributing to the
resonance $0.6597$ gives an
almost diverging contribution to the spectral determinant, thus hindering
its expansion to a
higher order.

In the cycle expansion the phases of the resonances are reproduced exactly
since they are again directly determined by lengths of orbits.
If $p$ is the shortest orbit length used in $d_i(z)$, the polynomial
can as well be written as a polynomial in $z^p$ thus allowing the zeros
to have the phases of the $p$th roots of unity.

As a final check on the physical meaning of our frozen eigenvalues with
moduli
smaller than unity as Frobenius-Perron resonances we have compared these
moduli  with rates of correlation decay. 
In a numerical experiment we investigated the decay of the correlator
$C(n) \!=\!
[\langle \rho(n) \rho(0) \rangle - \langle \rho(\infty) \rho(0)
\rangle]\cdot [\langle \rho(0) \rho(0) \rangle - \langle \rho(\infty)
\rho(0)
\rangle]^{-1}$.
Depending on the choice of $\rho(0)$ different long-time decays are
observable. We chose $\rho(0)$ as covering the regions where the
hyperbolic orbits relevant for a given resonance are situated.
Figure \ref{figure1}c illustrates the very good agreement between the
long-time decay
of $C(n)$ (dots) and the decay as predicted by the corresponding resonance
$0.81$ (full line).
Together with the resonances at $l_{max}=60$ table \ref{table2} displays 
the associated decay
factors by which $C(n)$ decreases over one timestep, obtained from a 
numerical fit. 
Again the agreement is convincing.

In conclusion, we have presented a method to determine Frobenius-Perron  
resonances and the associated phase-space structures, applicable to systems  
with mixed phase spaces. The acquired knowledge of phase space structures  
allows to check the
accuracy to which resonances are determined by the otherwise
independent approaches
of cycle expansion and correlation decay.

We are grateful to Shmuel Fishman for discussions initiating  as well as
accompanying this work.
Support by the Sonderforschungsbereich ``Unordnung und gro{\ss}e
Fluktuationen''
and by the grant GAAV No. A1048804 of the Czech Academy of Sciences is
thankfully
acknowledged. F. H. also thanks the Isaac Newton Institute for hospitality
during the workshop
``Supersymmetry and Trace Formulae'' in 1997 during which this work was
begun;
especially fruitful interactions with Ilya Goldscheid were made possible
there.

\onecolumn

\begin{table}
\[
\begin{array}{|r||r|r|r|} \hline
l_{max} = 40 & l_{max} = 50 & l_{max} = 60 & l_{max} = 70 \\ \hline\hline
0.8116 & 0.8205 & 0.8103 & 0.8076 \\ \hline
0.7457 & 0.7547 & 0.7470 & 0.7459 \\ \hline
-0.7432 & -0.7475 & -0.7510 & -0.7465 \\ \hline
\begin{array}{r} -0.0063\\ \pm {\rm i}\; 0.7431\end{array} &
\begin{array}{r} -0.0123\\ \pm {\rm i}\; 0.7515\end{array} &
\begin{array}{r} -0.0079\\ \pm {\rm i}\; 0.7517\end{array} &
\begin{array}{r} -0.0042\\ \pm {\rm i}\; 0.7414\end{array} \\ \hline
-0.6443 & -0.6188 & -0.6377 & -0.6347 \\ \hline
\end{array}
\]
\caption{Resonances appearing for $\tau=10$ at $l_{max} = 40, 50, 60$ and
$70$.}
\label{table1}
\end{table}

\begin{table}
\[
\begin{array}{|r||r|r|r|} \hline
l_{max} = 60 & 0.8103 & -0.7510 & 0.6597 \\ \hline\hline
n_{max} = 1 & 0.2185 \{1\} & - & - \\ \hline
n_{max} = 2 & 0.7070 \{2\} & - & 0.4969 \{1\} \\ \hline
n_{max} = 4 & 0.7664 \{3\} & -0.7483 \{4\} & - \\ \hline \hline
\mbox{decay of} \; C(n) & 0.8005 & 0.7697 & 0.6783 \\ \hline
\end{array}
\]
\caption{First row: resonances obtained from the truncated propagator
for $\tau = 10$ and $l_{max} = 60$. Below: corresponding results from cycle
expansion
up to order $n_{max}$. The total number of primitive orbits employed is
given in curly brackets.
Last row: associated decay factors by which $C(n)$ decreases over 
one timestep, obtained from numerical fit.}
\label{table2}
\end{table}

\end{document}